\documentclass[a4paper]{article}

\usepackage{natbib}
\usepackage{amssymb}
\usepackage{latexsym}
\usepackage{bussproofs}
\usepackage{url}
\usepackage{pxfonts}
\usepackage{microtype}%Improves spacing between letters and words, must be loaded after fonts

\usepackage{stackengine}
\stackMath

\newtheorem{definition}{Definition}

\newtheorem{lemma}{Lemma}
\newtheorem{theorem}{Theorem}
\newtheorem{corollary}{Corollary}

\begin{document}

\title{Normalisation and Subformula Property for a System of Classical Logic with Tarski's Rule}
\author{Nils K\"urbis}
\date{}
\maketitle

\begin{center}
Published in \emph{Archive for Mathematical Logic}  \\ 
\url{http://dx.doi.org/10.1007/s00153-021-00775-6}\bigskip
\end{center}

\begin{abstract}
\noindent This paper considers a formalisation of classical logic using general introduction rules and general elimination rules. It proposes a definition of `maximal formula', `segment' and `maximal segment' suitable to the system, and gives reduction procedures for them. It is then shown that deductions in the system convert into normal form, i.e. deductions that contain neither maximal formulas nor maximal segments, and that deductions in normal form satisfy the subformula property. Tarski's Rule is treated as a general introduction rule for implication. The general introduction rule for negation has a similar form. Maximal formulas with implication or negation as main operator require reduction procedures of a more intricate kind not present in normalisation for intuitionist logic.
\end{abstract} 


\section{Introduction} 
Deductions in normal form in Gentzen's formalisation of intuitionist logic in natural deduction \citep[186]{gentzenuntersuchungen} satisfy the \emph{subformula property}. Precise definitions will be given in the next section, but for this preliminary discussion, the following explanations of terminology suffice. A deduction is in normal form if it contains no maximal formula\footnote{We can distinguish occurrences of formulas, which in the present context occur for the most part in deductions, from a more abstract notion of formula which collects together formulas of the same shape or form, as it is customary to say. The former could also be referred to as formula types, the latter as their tokens. For brevity, by `formula' I will often mean an occurrence of a formula, but be explicit about the distinction where this aids understanding. There are also schematic formulas and their instances, which may or may not be formulas of the same shape, and we can distinguish the general statement of a rule of inference from its application in a deduction: the former is made in terms of schematic formulas and specifies the common form of all its instances, the latter have formula occurrences as their premises and conclusions and are used in the construction of deductions. I will thus speak of rules as well as of their applications, but for brevities sake by `rule' I will often mean an application of a rule. This clarification and the ensuing greater precision in the use of terminology in this paper was added at the request of a referee for this journal, to whom I thank for the helpful comments on this paper. A second referee also made valuable comments on the previous version of this paper and pointed me to Michel Parigot's work, which is discussed in section 3. This referee also suggested a comparison of the results reported here with Seldin's \citep{seldin}, but this must wait for another occasion. I would like to thank Andrzej Indrzejczak, Michał Zawidski and Yaroslav Petrukhin for comments on this paper, and Peter Milne for his system, discussions and support. During the preparation of the final version of this paper I was an Alexander von Humboldt research fellow at the University of Bochum, to whom also many thanks are due.}, that is a formula that is the conclusion of an introduction rule and the major premise of an elimination rule for its main connective. A deduction has the subformula property if any formula that occurs on it is a subformula of either an undischarged assumption or of the conclusion. The normalisation theorem for intuitionist logic states that any deduction can be brought into normal form, and so for any deduction of $A$ from assumptions $\Gamma$ in intuitionist logic, there is a deduction of $A$ from some of the assumptions in $\Gamma$ that satisfies the subformula property. This was shown by Prawitz \citep[Ch IV]{prawitznaturaldeduction}.\footnote{Von Plato has edited previously unpublished material of Gentzen's that shows that he had also proved these results for intuitionist logic (See \citep{platogentzennormalisation} and \citep{platocellar}).}

Systems of natural deduction for classical propositional logic in which any deduction converts into one with the subformula property have proved to be more elusive. In general, deductive systems for classical logic with this property relinquish some of the features of natural deduction for intuitionist logic. The obvious route is to abandon single conclusions and adopt Gentzen's sequent calculus. Deductions in normal form in Prawitz's system of natural deduction for classical logic, for instance, only satisfy a restricted version of the subformula property: allowance must be made for assumptions of the form $\neg A$ that are discharged by classical \emph{reductio ad absurdum} and formulas $\bot$ concluded from them \citep[42]{prawitznaturaldeduction}. 

The main result of this paper is a proof that in a system of natural deduction for classical propositional logic \textbf{C} put forward by Milne \citep{milneinversion}, every deduction converts into one in normal form, and every deduction in normal form satisfies the subformula property \emph{without} any restrictions. The system has some rather original features. It couples the well-known \emph{general elimination rules} for $\lor$, $\land$, $\supset$\footnote{See \citep{platogeneral}. Such rules were also considered by Prawitz in \citep{prawitzmeaningandcompleteness}.} with introduction rules that are unusual in that, rather than permitting the conclusion of a formula with the connective they govern as main operator, they permit the discharge of such a formula. These were called `general introduction rules' by Negri and von Plato, who formalised a system of intuitionist logic with such rules \citep[217]{negriplatostructural}. While Negri and von Plato use $\bot$ as primitive, and define negation in terms of it, Milne has general introduction and elimination rules for a primitive negation operator. Milne's general introduction rules for implication are also unexpected and original. Negri and von Plato's system is the subject of an independent investigation \citep{kurbisgenintroint}, in which normalisation and subformula property are proved for this system, too, and general introduction and elimination rules for identity are also considered. 

Milne's system is closely related to a formalisation of classical logic by Michel Parigot \citep{parigotfreededuction}. The main difference is that Parigot's system is formalised in sequent calculus, but translating one framework into the other, they are, \emph{mutatis mutandis}, the same. A brief section of this paper compares the two systems. 

Milne observes that, while adding rules for the universal quantifier to his system cannot produce one in which any deduction can be transformed into one that satisfies the subformula property, this is not so for the existential quantifier. The penultimate section of the present paper sheds some further light on this result: if the universal quantifier is added, the reduction procedures for maximal formulas with negation as main operator, and one of the reduction procedures for maximal formulas with implication as main operator, are no longer applicable. Thus deductions in the system do not normalise. The problem does not arise when the existential quantifier is added: any deduction in \textbf{C} plus $\exists$ converts into one in normal form, and deductions in normal form have the subformula property. 

Finally, the conclusion considers transposing general introduction rules back into the more usual form of introduction rules. 

Milne gives a model-theoretical and non-constructive proof that in his system, if, whenever all of $\Gamma$ are true, then $A$ is true, then there is a deduction of $A$ from $\Gamma$ that has the subformula property \citep{milnesubformula}. By soundness it follows that for any deduction in his system, there is one that satisfies the subformula property. Here this result is proved proof-theoretically and constructively as a consequence of the normalisation theorem for the system. 

The results of this paper are of some philosophical interest. Normalisation and subformula property of deductions in normal form are commonly regarded as necessary conditions for harmony.\footnote{Dummett provides the most extended discussion of harmony and its meaning-theoretical foundations \cite{dummettLBM}. Prawitz made many contributions to the surrounding debate, for instance \citep{prawitzdummett} and \citep{prawitzdummett2}.} Harmony consists in a certain balance in the rules governing a logical constant: an elimination rule for a logical connective should not license the deduction of more conclusions from a formula with it as main operator than are justified by its introduction rules. Prawitz's formalisation of classical logic of \citep{prawitznaturaldeduction}, for instance, lacks the subformula property for normal deductions and its rules for negation are not considered to be harmonious. The disappearance of a negation from the assumptions and conclusions of a deduction in those cases where the strict subformula property is violated establish that too much has been derived from the discharged assumptions: Prawitz's rules for classical negation license the deduction of more conclusions from formulas of the form $\neg A$ than are justified by the introduction rule governing this connective. A more detailed discussion of the philosophical implications of the results presented here goes beyond the purposes of this paper.

\section{A System of Classical Propositional Logic} 
The definition of the language of \textbf{C} is standard. 

\begin{definition}[Connective, Atomic Formula, Degree of a Formula]\label{D1}
\normalfont $\neg$, $\supset$, $\land$ and $\lor$ are the \emph{connectives}. An \emph{atomic formula} is one that contains no connective. The \emph{degree} of a formula is the number of connectives occurring in it. 
\end{definition}

\noindent Occasionally, the discussion mentions $\bot$. Being a connective, it is not an atomic formula, but a formula of degree 1.

Deductions in \textbf{C} have the familiar tree shape, with the (discharged or undischarged) assumptions at the top-most nodes or leaves and the conclusion at the bottom-most node or root. The conclusion of a deduction is said to \emph{depend on} the undischarged assumptions of the deduction. Similar terminology is applied to subdeductions of deductions. 

Every assumption in a deduction belongs to an \emph{assumption class}, marked by a natural number, different numbers for different assumption classes. Formula occurrences of different types\footnote{See footnote 1.} must belong to different assumption classes. Formula occurrences of the same type may, but do not have to, belong to the same assumption class. Discharge of assumptions is marked by a square bracket around the formula: $[A]^i$, $i$ being the assumption class to which $A$ belongs, with the same label of the assumption class also occurring at the application of the rule at which the assumption is discharged. Assumption classes are chosen in such a way that if one assumption of an assumption class is discharged by an application of a rule, then it discharges all assumptions of said assumption class. Empty assumption classes are permitted: they are used in vacuous discharge, when a rule that allows for the discharge of assumptions is applied with no assumptions being discharged. 

Upper case Greek letters $\Sigma$, $\Pi$, $\Xi$, possibly with subscripts or superscripts, denote deductions. In general some of the assumptions and the conclusion of the deduction are mentioned explicitly at the top and bottom of $\Sigma$, $\Pi$, $\Xi$. Using the same designation more than once to denote subdeductions of a deduction means that these subdeductions are exact duplicates of each other apart from, possibly, the labels for assumption classes: the deductions have the same structure and at every node formulas of the same type are premises and conclusions of applications of the same rules. 

\begin{definition}[Deduction in \textbf{C}]\label{D2}\ \\
\normalfont (i) The formula occurrence $A^n$, where $n$ marks an assumption class, is a deduction in \textbf{C} of $A$ from the undischarged assumption $A$. 

\noindent (ii) If $\Sigma$, $\Pi$, $\Xi$ are deductions in \textbf{C}, then following are deductions of $C$ in \textbf{C} from the undischarged assumptions in $\Sigma$, $\Pi$, $\Xi$ apart from those in the assumption classes $i$ and $j$, which are discharged: 

\begin{center} 
\bottomAlignProof
\AxiomC{$\Sigma$}
\noLine
\UnaryInfC{$A$}
\AxiomC{$\Pi$}
\noLine
\UnaryInfC{$B$}
\AxiomC{$[A\land B]^i$}
\noLine
\UnaryInfC{$\Xi$}
\noLine
\UnaryInfC{$C$}
\RightLabel{$_{\land I \ i}$}
\TrinaryInfC{$C$}
\DisplayProof\qquad
\bottomAlignProof
\AxiomC{$\Sigma$}
\noLine
\UnaryInfC{$A\land B$}
\AxiomC{$[A]^i \ [B]^j$}
\noLine
\UnaryInfC{$\Pi$}
\noLine
\UnaryInfC{$C$}
\RightLabel{$_{\land E \ i, j}$}
\BinaryInfC{$C$}
\DisplayProof

\bigskip 

\AxiomC{}
\noLine
\UnaryInfC{$\Sigma$}
\noLine
\UnaryInfC{$A$}
\AxiomC{$[A\lor B]^i$}
\noLine
\UnaryInfC{$\Pi$}
\noLine
\UnaryInfC{$C$}
\RightLabel{$_{\lor I \ i}$}
\BinaryInfC{$C$}
\DisplayProof\quad
\AxiomC{}
\noLine
\UnaryInfC{$\Sigma$}
\noLine
\UnaryInfC{$B$}
\AxiomC{$[A\lor B]^i$}
\noLine
\UnaryInfC{$\Pi$}
\noLine
\UnaryInfC{$C$}
\RightLabel{$_{\lor I \ i}$}
\BinaryInfC{$C$}
\DisplayProof\qquad
\AxiomC{}
\noLine
\UnaryInfC{$\Sigma$}
\noLine
\UnaryInfC{$A\lor B$}
\AxiomC{$[A]^i$}
\noLine
\UnaryInfC{$\Pi$}
\noLine
\UnaryInfC{$C$}
\AxiomC{$[B]^j$}
\noLine
\UnaryInfC{$\Xi$}
\noLine
\UnaryInfC{$C$}
\RightLabel{$_{\lor E \ i, j}$}
\TrinaryInfC{$C$}
\DisplayProof

\bigskip

\AxiomC{$\Pi$}
\noLine
\UnaryInfC{$B$}
\AxiomC{$[A\supset B]^i$}
\noLine
\UnaryInfC{$\Sigma$}
\noLine
\UnaryInfC{$C$}
\RightLabel{$_{\supset I \ i}$}
\BinaryInfC{$C$}
\DisplayProof\qquad
\AxiomC{$[A]^i$}
\noLine
\UnaryInfC{$\Pi$}
\noLine
\UnaryInfC{$C$}
\AxiomC{$[A\supset B]^j$}
\noLine
\UnaryInfC{$\Sigma$}
\noLine
\UnaryInfC{$C$}
\RightLabel{$_{TR \ i, j}$}
\BinaryInfC{$C$}
\DisplayProof\qquad
\AxiomC{$\Pi$}
\noLine
\UnaryInfC{$A\supset B$}
\AxiomC{}
\noLine
\UnaryInfC{$\Sigma$}
\noLine
\UnaryInfC{$A$}
\AxiomC{$[B]^i$}
\noLine
\UnaryInfC{$\Xi$}
\noLine
\UnaryInfC{$C$}
\RightLabel{$_{\supset E \ i}$}
\TrinaryInfC{$C$}
\DisplayProof\bigskip
 
\AxiomC{$[A]^i$}
\noLine
\UnaryInfC{$\Pi$}
\noLine
\UnaryInfC{$C$}
\AxiomC{$[\neg A]^j$}
\noLine
\UnaryInfC{$\Sigma$}
\noLine
\UnaryInfC{$C$}
\RightLabel{$_{\neg I \ i, j}$}
\BinaryInfC{$C$}
\DisplayProof\qquad
\AxiomC{$\Pi$}
\noLine
\UnaryInfC{$\neg A$}
\AxiomC{$\Sigma$}
\noLine
\UnaryInfC{$A$}
\RightLabel{$_{\neg E}$}
\BinaryInfC{$C$}
\DisplayProof
\end{center}

\noindent (iii) Nothing else is a deduction in \textbf{C}.
\end{definition} 

\noindent In writing deductions we may suppress the label indicating the rule applied, but the labels indicating discharge must always be present. 

A few words on the rules for implication will be in order. In Negri and von Plato's intuitionist system with general introduction and elimination rules, the introduction rule for $\supset$ permits the discharge of an assumption $A$ above its first premise $B$. Thus letting $C$ be $A\supset B$ in their rule derives the usual introduction rule for $\supset$, which licenses the deduction to $A\supset B$ from a deduction of $B$ from $A$. $TR$ is named after Tarski: Milne calls it \emph{Tarski's Rule}. Following the patterns of the other rules, it is treated as a second general introduction rule for $\supset$. The usual introduction rule for $\supset$ is derived in \textbf{C} in the following way:  

\begin{prooftree}
\AxiomC{$[A]^2$}
\noLine
\UnaryInfC{$\Pi$}
\noLine
\UnaryInfC{$B$}
\AxiomC{$[A\supset B]^1$}
\RightLabel{$_{1 \ \supset I}$} 
\BinaryInfC{$A\supset B$}
\AxiomC{$[A\supset B]^2$}
\RightLabel{$_{2 \ TR}$} 
\BinaryInfC{$A\supset B$}
\end{prooftree} 

\noindent The usual negation introduction rule, \emph{reductio ad absurdum}, licenses the deduction of $\neg A$ from a deduction of a contradiction from $A$. It is derived  like this in \textbf{C}: 

\begin{prooftree}
\AxiomC{$[A]^1$}
\noLine
\UnaryInfC{$\Pi$}
\noLine
\UnaryInfC{$\neg B$} 
\AxiomC{$[A]^1$}
\noLine
\UnaryInfC{$\Sigma$}
\noLine
\UnaryInfC{$B$} 
\RightLabel{$_{\neg E}$} 
\BinaryInfC{$\neg A$}
\AxiomC{$[\neg A]^1$}
\RightLabel{$_{1 \ \neg I}$}
\BinaryInfC{$\neg A$}
\end{prooftree} 

\noindent Classical \emph{reductio ad absurdum}, which licenses the deduction of $A$ from a deduction of a contradiction from $\neg A$, is derived by interchanging $A$ and $\neg A$. Conversely, Milne's general introduction rules are clearly valid rules of inference of classical logic. Thus we have a system of classical propositional logic.  

When added to the system, $\bot$ is governed by a sole elimination rule, \emph{ex falso quodlibet}. It licenses the deduction of any formula from $\bot$. The rule may be restricted to atomic conclusions. It is the limiting case of a general elimination rule where there are no side-deductions with discharged assumptions corresponding to subformulas of the major premise above them. Having no introduction rule, it is trivially governed only by rules of the same form as the other connectives. 
 
Notice that $\neg E$ has the form of a general elimination rule. This can be seen by treating $\neg A$ as $A\supset \bot$ and replacing $\bot$ for $B$ in $\supset E$. As everything follows from $\bot$, a side-deduction showing that the conclusion of the application of the rule follows from $\bot$ is redundant. 

For recording from which assumptions a conclusion has been derived, it suffices to list the assumption classes to which the undischarged assumptions of the deduction belong. This will be a multiset. If a set is preferred, delete multiple occurrences of assumption classes of formulas of the same type. We can write $\Gamma\vdash_\mathbf{C} A$ if there is a deduction of (the formula occurrence) $A$ from (occurrences of) some of the formulas in $\Gamma$ in \textbf{C}. 

The premise $A$ of $\supset E$ and $\neg A$ and $C$ in all four elimination rules are normally called the minor premises, but in the current system it is useful to have terminology that allows to distinguish them. 

\begin{definition}[Terminology for Premises and Discharged Assumptions]\label{majorminoretc}\ \\
\normalfont (i) In applications of the elimination rules, formula occurrences taking the places of $A\land B$, $A\supset B$, $A\lor B$ and $\neg A$ to the very left above the line are the \emph{major premises}; formula occurrences taking the places of $C$ are the \emph{arbitrary premises}, and a formula occurrence taking the place of $A$ in an application of $\supset E$ and in $\neg E$  is the \emph{minor premise} of those rules. 

\noindent (ii) In applications of $\land I$, $\lor I$, $\supset I$, formula occurrences taking the places of $A$ and $B$ to the very left above the line are the \emph{specific premises} (both in case of $\land I$), and those taking the place of $C$ to their right are also called the \emph{arbitrary premises}; formula occurrences taking the places of the discharged assumptions $A\land B$, $A\lor B$, $A\supset B$ and $\neg A$ are the \emph{major assumptions discharged by} applications of $\land I$, $\lor I$, $\supset I$, $TR$ and $\neg I$, and those taking the places of the discharged assumptions $A$ in $\neg I$ and $TR$ are the \emph{minor assumptions discharged by} applications of those rules.  
\end{definition}

\noindent Like $\bot E$, $\neg I$ has no arbitrary premises.

Vacuous discharge happens when in an application of a rule that permits discharge no formula is in fact discharged. In systems of natural deduction with the rules of Gentzen and Prawitz, applications of rules with vacuous discharge above arbitrary premises are removed from deductions by what is often called \emph{simplification conversions}\footnote{See, e.g., \citep[181]{troelstraschwichtenberg}.} The procedure is obvious, and I will give no details here. In \textbf{C}, all vacuous discharge happens above arbitrary premises, and there it is clearly superfluous: instead of applying a rule and discharging vacuously, we might as well go on with the deduction straight from the arbitrary premise. In \textbf{C} there is no need for vacuous discharge at all, and in the following, I will assume that any deduction is cleaned up so as to contain no vacuous discharge above arbitrary premises; in particular, I will assume that this is done should vacuous discharge above an arbitrary premise arise as a result of the conversions of deductions to be given in section 4. In $\land E$, it is of course often necessary to make use of the option of discharging only one assumption.

\section{Comparison with Parigot's Free Deduction}
The feature of Milne's system that all conclusions of rules are identical to an arbitrary premise and hence no logical constants appear below the inference line in any of the general statements of its rules has a close parallel in Michel Parigot's system of free deduction \citep{parigotfreededuction}. This is a system of multiple conclusion sequent calculus for classical logic where it is also the case that no formulas with logical constants appear in the conclusions of the general statement of the rules: 

\bigskip

\noindent Axiom\bigskip

$A\vdash A$

\bigskip

\noindent Conjunction\bigskip

\AxiomC{$\Gamma, A\land B\vdash \Delta$}
\AxiomC{$\Pi\vdash A, \Sigma$}
\AxiomC{$\Pi'\vdash B, \Sigma'$}
\LeftLabel{$L\land$:} 
\TrinaryInfC{$\Gamma, \Pi, \Pi'\vdash \Delta, \Sigma, \Sigma'$}
\DisplayProof

\bigskip

\AxiomC{$\Gamma\vdash A\land B, \Delta$}
\AxiomC{$\Pi, A\vdash \Sigma$}
\LeftLabel{$R\land$:}
\BinaryInfC{$\Gamma, \Pi\vdash \Delta, \Sigma$}
\DisplayProof\qquad
\AxiomC{$\Gamma\vdash A\land B, \Delta$}
\AxiomC{$\Pi, B\vdash \Sigma$}
\BinaryInfC{$\Gamma, \Pi\vdash \Delta, \Sigma$}
\DisplayProof

\bigskip

\noindent Disjunction\bigskip

\AxiomC{$\Gamma, A\lor B\vdash \Delta$}
\AxiomC{$\Pi\vdash A, \Sigma$}
\LeftLabel{$L\lor$:}
\BinaryInfC{$\Gamma, \Pi\vdash\Delta, \Sigma$} 
\DisplayProof\qquad
\AxiomC{$\Gamma, A\lor B\vdash \Delta$}
\AxiomC{$\Pi\vdash B, \Sigma$}
\BinaryInfC{$\Gamma, \Pi\vdash\Delta, \Sigma$}
\DisplayProof

\bigskip

\AxiomC{$\Gamma\vdash A\lor B, \Delta$}
\AxiomC{$\Pi, A\vdash \Sigma$}
\AxiomC{$\Pi', B\vdash \Sigma'$}
\LeftLabel{$R\lor$:} 
\TrinaryInfC{$\Gamma, \Pi, \Pi'\vdash \Delta, \Sigma, \Sigma'$}
\DisplayProof

\bigskip

\noindent Negation\bigskip

\AxiomC{$\Gamma, \neg A\vdash \Delta$} 
\AxiomC{$\Pi, A\vdash \Sigma$}
\LeftLabel{$L\neg$:}
\BinaryInfC{$\Gamma, \Pi\vdash\Delta, \Sigma$} 
\DisplayProof\qquad
\AxiomC{$\Gamma \vdash \neg A, \Delta$} 
\AxiomC{$\Pi\vdash A, \Sigma$}
\LeftLabel{$R\neg$:}
\BinaryInfC{$\Gamma, \Pi\vdash\Delta, \Sigma$} 
\DisplayProof

\bigskip

\noindent Implication\bigskip

\AxiomC{$\Gamma, A\supset B\vdash \Delta$}
\AxiomC{$\Pi, A\vdash\Sigma$}
\LeftLabel{$L\supset$:}
\BinaryInfC{$\Gamma, \Pi\vdash\Delta, \Sigma$} 
\DisplayProof\qquad
\AxiomC{$\Gamma, A\supset B\vdash \Delta$}
\AxiomC{$\Pi\vdash B, \Sigma$}
\BinaryInfC{$\Gamma, \Pi\vdash\Delta, \Sigma$} 
\DisplayProof

\bigskip

\AxiomC{$\Gamma\vdash A\supset B, \Delta$} 
\AxiomC{$\Pi\vdash A, \Sigma$}
\AxiomC{$\Pi', B\vdash \Sigma'$}
\LeftLabel{$R\supset$:}
\TrinaryInfC{$\Gamma, \Pi, \Pi'\vdash \Delta, \Sigma, \Sigma'$}
\DisplayProof

\bigskip

\noindent $\Gamma, \Delta, \Pi, \Sigma$ are sets of formulas. Weakening and contraction may be added either in their usual form as explicit rules or in an implicit form through the conventions that active formulas need not occur in a premise and that the conclusions of rules are contracted. 

Parigot explains how to obtain common systems of sequent calculus and natural deduction for classical logic from free deduction. In both cases, some of the premises of the rules of free deduction are replaced by axioms and then suppressed \citep[367f]{parigotfreededuction}. Parigot gives the examples of a common formulation of the classical sequent calculus\footnote{With explicit rules for weakening and contraction on the left and right, it is the calculus called \textbf{G1c} with a primitive negation operator and without $\bot$ \citep[61]{troelstraschwichtenberg}, and with alternative rules for implication: Parigot's system has two right rules for $\supset$: in one rule $\Pi\vdash A\supset B, \Sigma$ is inferred from $\Pi, A\vdash \Sigma$, in the other from $\Pi\vdash B,\Sigma$.} and a system of natural deduction with multiple conclusions. 

Milne's system can be obtained from Parigot's by using replacing its right conjunction rules by a variant \citep[364]{parigotfreededuction}:\bigskip 

\AxiomC{$\Gamma\vdash A\land B, \Delta$}
\AxiomC{$\Pi, A, B\vdash \Sigma$}
\LeftLabel{$R\land'$:} 
\BinaryInfC{$\Gamma, \Pi\vdash \Delta, \Sigma$}
\DisplayProof

\bigskip

\noindent Instead of replacing premises by axioms and suppressing them, we simply restrict the succedents of sequents of free deduction to single conclusions in the most obvious way. Right rules turn into general elimination rules. Left rules turn into general introduction rules. Not carrying an active formula of a rule of free deduction to the left of $\vdash$ into the conclusion corresponds to the discharge of that formula; active formulas to the right of $\vdash$ turn into premises of the rules:\bigskip 

$\land I$: take $L\land$, let $\Delta=\{C\}$ and let $\Sigma=\Sigma'=\varnothing$. 

$\land E$: take $R\land'$, let $\Sigma=\{C\}$ and let $\Delta=\varnothing$. 

$\lor I$: take $L\lor$, let $\Delta=\{C\}$ and let $\Sigma=\varnothing$. 

$\lor E$: take $R\lor$, let $\Delta=\varnothing$ and let $\Sigma=\Sigma'=\{C\}$.

$\neg I$: take $L\neg$, let $\Delta=\Sigma=\{C\}$. 

$\neg E$: take $L\neg$, let $\Delta=\Sigma=\varnothing$ in the premise and $\{C\}$ in the conclusion.

$TR$: take the first $L\supset$, let $\Delta=\Sigma=\{C\}$ 

$\supset I$: take the second $L\supset$ and let $\Delta=\{C\}$ and let $\Sigma=\varnothing$.  

$\supset E$: take $R\supset$ and let $\Delta=\Sigma=\varnothing$ and let $\Sigma'=\{C\}$. 

\bigskip \noindent The result is Milne's system when written in natural deduction in sequent calculus style. 

At one point the correspondence is not quite so perfect: in $\neg E$, the succedents of the premises and the conclusion are treated differently. But this is a minor blemish. It could be remedied by adding $\bot$ to the system and treating $\neg$ as defined in terms of it and $\supset$.   

A closer comparison of the proof-theoretic properties of Milne's system with Parigot's free deduction and other systems of classical logic found in the literature merits its own investigation is left for another occasion.

\section{Normalisation for C}
We begin by adjusting the common notion of a maximal formula to deductions in \textbf{C}: 

\begin{definition}[Maximal Formula]\label{maximalformula}
\normalfont A \emph{maximal formula} in a deduction in \textbf{C} with main operator $\ast$ is an occurrence of a formula $A\ast B$  or $\ast \ A$ (i.e. $\ast$ is $\neg$) that is the major premise of an application of $\ast E$ and the major assumption discharged by an application of $\ast I$ or $TR$ in case $\ast$ is $\supset$. 
\end{definition} 

\noindent Maximal formulas are removed from deductions by procedures to be given shortly. But first an observation about negation. 

It would be possible to restrict the conclusion of $\neg E$ to formulas not having $\supset$, $\land$ or $\lor$ as main connectives. Similar constructions that show that $\bot E$ can be restricted to atomic conclusions are easily adapted to $\neg E$ of \textbf{C}. We cannot, however, restrict the conclusion to atomic formulas nor to literals: they sometimes need to be of the form $\neg C$, where $C$ may be a complex formula. 

An attempt to avoid deriving $\neg C$ by $\neg E$ while also avoiding the creation of a maximal formula would need to appeal to $\neg I$ where the major assumption discharged by the application of the rule is also the right arbitrary premise and of the same shape as the conclusion. Hence we would need to derive $\neg C$ from $C$ to gain the left arbitrary premise using the premises $\neg A$ and $A$ of the application of $\neg E$ that concluded $\neg C$. The only way to do so would be to derive $\neg C$ from $A$ and $\neg A$, which was to be avoided. 

$\neg E$ will therefore be treated like all the other elimination rules and it will be shown that its conclusion need never be the major premise of an elimination rule. 

Applications of general introduction and elimination rules except $\neg E$ require deductions of arbitrary premises $C$ which also provide the conclusion of the application of the rule. They form part of a sequence of formula occurrences of the same shape\footnote{See footnote 1.}:

\begin{definition}[Segment]\label{segment}
\normalfont A \emph{segment} is a sequence of formula occurrences $C_1\ldots C_n$ in a deduction such that either

\noindent (i) $n>1$ and for all $i<n$, $C_i$ is an arbitrary premise of an application of a rule and $C_{i+1}$ is its conclusion, and $C_n$ is not an arbitrary premise of an application of a rule; or 

\noindent (ii) $n\geq 1$ and $C_1$ is the conclusion of $\neg E$ and for all $i<n$, $C_i$ is an arbitrary premise of an application of a rule and $C_{i+1}$ is its conclusion, and $C_n$ is not an arbitrary premise of an application of a rule.
\end{definition} 

\noindent The \emph{length} of a segment is the number of formula occurrences of which it consists, its \emph{degree} the degree of any such formula. As $C_1\ldots C_n$ are all of the same shape, I will speak of the formula (as a type) constituting the segment. 

Every conclusion of $\neg E$ forms part of a segment of length 1. The motivation for classifying conclusions of $\neg E$ as giving rise to segments is that this rule is like a crossing between $\supset E$ and $\bot E$: it has a minor premise, like $\supset E$, but lacks arbitrary premises, like $\bot E$.  

Due to the ban on vacuous discharge above arbitrary premises, in all other cases the first formula of a segment is an arbitrary premise discharged by an introduction rule the conclusion of which is the second formula of the segment. 

The major, minor and specific premises of rules are either assumptions or the last formulas of segments. 

\begin{definition}[Maximal Segment]\label{maximalsegment} 
\normalfont A \emph{maximal segment} is a segment the last formula of which is the major premise of an elimination rule. 
\end{definition} 

\noindent Maximal segments are removed from deductions by procedures to be given shortly. It will be a consequence of their successive application that maximal segments the first formula of which is the conclusion of $\neg E$ are removed from deductions. 

\begin{definition}[Normal Form]\label{normalform}
\normalfont A deduction is in \emph{normal form} if it contains neither maximal formulas nor maximal segments. 
\end{definition}

\noindent It will be shown that any deduction can be brought into normal form by successive and systematic application of the procedures to be given now, which remove maximal segments and formulas from deductions. I will call the deduction to which such a procedure is applied the \emph{initial} deduction and the result the \emph{reduced} deduction. 

First we need to take care of a minor issue. In the initial deduction, more assumptions of the form $A\ast B$ or $\ast \ A$ than the maximal formula to be removed by the procedure may be discharged above the arbitrary premise of the introduction rule. There are several ways of dealing with this issue. A fuller discussion of the options is reserved for the companion piece to the present paper which covers intuitionist logic \citep{kurbisgenintroint}. Here I will simply adopt the easiest solution to the problem, which consists in the observation that any application of a general introduction rule can be transformed into one that discharges exactly one major assumption. An advantage besides simplicity is that the normalisation theorem for \textbf{C} with rules so restricted transposes directly to a system of classical logic with the more common introduction rules in the style of Gentzen and Prawitz. This will be established in the conclusion to the present paper. 

Suppose, for instance, one wanted to discharge $n$ formula occurrences of the type $A\lor B$ by an application of $\lor I$: 

\begin{prooftree}
\AxiomC{$\Sigma$}
\noLine
\UnaryInfC{$A$}
\AxiomC{$\underbrace{[A\lor B]^i, [A\lor B]^i \ldots [A\lor B]^i}$}
\noLine
\UnaryInfC{$\Pi$}
\noLine
\UnaryInfC{$C$}
\RightLabel{$_i$}
\BinaryInfC{$C$}
\end{prooftree}

\noindent Then instead of making this one application of $\lor I$, one can apply it $n$ times: 

\begin{prooftree}
\AxiomC{$\Sigma$}
\noLine
\UnaryInfC{$A$}
\AxiomC{$\Sigma$}
\noLine
\UnaryInfC{$A$}
\AxiomC{$\Sigma$}
\noLine
\UnaryInfC{$A$}
\AxiomC{$\underbrace{[A\lor B]^1, [A\lor B]^2 \ldots [A\lor B]^n}$}
\noLine
\UnaryInfC{$\Pi$}
\noLine
\UnaryInfC{$C$}
\RightLabel{$_1$}
\BinaryInfC{$C$}
\RightLabel{$_2$}
\BinaryInfC{$C$}
\noLine
\UnaryInfC{$\vdots$}
\noLine
\UnaryInfC{$C$}
\RightLabel{$_n$}
\BinaryInfC{$C$}
\end{prooftree} 

\noindent The cases for the other connectives are similar. 

There are now two options: either make the restriction part of the construction of deductions, or first transform a given deduction into one that satisfies the restriction before the reduction procedures are applied. Either option works, as the system with the unrestricted introduction rules and the system with their restricted versions are evidently equivalent. Obviously any application of a restricted introduction rule is also a correct application of the unrestricted version, and the converse holds in virtue of the following: 

\begin{lemma}\label{uniquedischarge}
Any deduction can be transformed into one in which every application of a general introduction rules discharges exactly one major assumption.
\end{lemma} 

\noindent \emph{Proof.} By the ban on vacuous discharge above arbitrary premises, the transformations indicated above and an induction over a suitable measure of the complexity of deductions, e.g. the number of applications of introduction rules discharging multiple formula occurrences of highest degree in a deduction. Take such an application such that no other such application stands above it in the deduction. Applying the transformation reduces the measure.\bigskip

\noindent Moreover, any sequence of applications of introduction rules as in the example above can be collapsed into one application, so one could, after maximal formulas have been removed from a deduction, also simplify it again in that respect. 

As it lends itself to a very straightforward normalisation proof, unless otherwise stated, in the following I assume what I call the \emph{unique discharge convention} on introduction rules: every application of an introduction rule for $\ast$ discharges exactly one formula occurrence of the form $A\ast B$ or $\ast \ A$. 

Now for the procedures to remove maximal segments and formulas from deductions. Concerning the procedures to remove maximal formulas of the form $A\supset B$ and $\neg A$, observe once more that all discharge happens above arbitrary premises, and so they can be used to conclude any formula. This observation is used to ensure that assumptions that may have been discharged in the initial deduction are also discharged in the reduced deduction: we will reuse certain applications of rules that discharge assumptions in the initial deduction with new arbitrary premises and conclusions in the reduced deduction for the purpose of ensuring the conclusion of the reduced deduction does not depend on more formulas than the conclusion of the initial deduction. 

\bigskip 

\noindent \emph{Reduction Procedures for Maximal Formulas} 

\noindent Maximal formulas are removed from deductions by applying the following \emph{reduction procedures for maximal formulas}, where $\Pi, \Sigma$ above $[A], [B]$ indicate that these deductions are used to conclude each formula occurrence in the assumption class to which $A, B$ belong (assumption class markers are deleted):\bigskip 

\noindent 1. The maximal formula is of the form $A\land B$. Convert the deduction on the left into the deduction on the right: 

\begin{center}
\AxiomC{$\Sigma_1$}
\noLine
\UnaryInfC{$A$}
\AxiomC{$\Sigma_2$}
\noLine
\UnaryInfC{$B$}
\AxiomC{$[A\land B]^k$}
\AxiomC{$[A]^i \ [B]^j$}
\noLine
\UnaryInfC{$\Pi_1$}
\noLine
\UnaryInfC{$C$}
\RightLabel{$_{i, j}$}
\BinaryInfC{$C$}
\noLine
\UnaryInfC{$\Pi_2$}
\noLine
\UnaryInfC{$D$}
\RightLabel{$_k$}
\TrinaryInfC{$D$}
\DisplayProof\quad$\leadsto$\quad
\alwaysNoLine
\AxiomC{$\mathbin{\stackon[6pt]{[A]}{\Sigma_1}} \ \mathbin{\stackon[6pt]{[B]}{\Sigma_2}}$}
\UnaryInfC{$\Pi_1$}
\UnaryInfC{$C$}
\UnaryInfC{$\Pi_2$}
\UnaryInfC{$D$}
\DisplayProof
\end{center}

\noindent 2. The maximal formula is of the form $A\lor B$. Convert the deduction on the left into the deduction on the right: 

\begin{center} 
\AxiomC{$\Sigma_1$}
\noLine
\UnaryInfC{$A$}
\AxiomC{$[A\lor B]^k$}
\AxiomC{$[A]^i$}
\noLine
\UnaryInfC{$\Pi_1$}
\noLine
\UnaryInfC{$C$}
\AxiomC{$[B]^j$}
\noLine
\UnaryInfC{$\Pi_2$}
\noLine
\UnaryInfC{$C$}
\RightLabel{$_{i, j}$}
\TrinaryInfC{$C$}
\noLine
\UnaryInfC{$\Pi_3$}
\noLine
\UnaryInfC{$D$}
\RightLabel{$_k$}
\BinaryInfC{$D$}
\DisplayProof\quad$\leadsto$\quad
\alwaysNoLine
\AxiomC{$\Sigma_1$}
\UnaryInfC{$[A]$}
\UnaryInfC{$\Pi_1$}
\UnaryInfC{$C$}
\UnaryInfC{$\Pi_3$}
\UnaryInfC{$D$}
\DisplayProof
\end{center} 

\noindent Similarly for the case where the premise of $\lor I$ is $B$ concluded by $\Sigma_2$.\bigskip

\noindent 3. The major premise of $\supset E$ is discharged by $\supset I$. Convert the deduction on the left into the deduction on the right: 

\begin{center} 
\AxiomC{$\Sigma$}
\noLine
\UnaryInfC{$B$}
\AxiomC{$[A\supset B]^i$}
\AxiomC{$\Pi_1$}
\noLine
\UnaryInfC{$A$}
\AxiomC{$[B]^j$}
\noLine
\UnaryInfC{$\Pi_2$}
\noLine
\UnaryInfC{$C$}
\RightLabel{$_j$}
\TrinaryInfC{$C$}
\noLine
\UnaryInfC{$\Xi$}
\noLine
\UnaryInfC{$D$}
\RightLabel{$_i$}
\BinaryInfC{$D$}
\DisplayProof\quad$\leadsto$\quad
\alwaysNoLine
\AxiomC{$\Sigma$}
\UnaryInfC{$[B]$}
\UnaryInfC{$\Pi_2$}
\UnaryInfC{$C$}
\UnaryInfC{$\Xi$}
\UnaryInfC{$D$}
\DisplayProof
\end{center} 

\noindent 4. The major premise of $\supset E$ is discharged by $TR$. Convert the deduction on the left into the deduction on the right, where $\Xi^\ast$ is constituted by applications of rules in $\Xi$ that discharge assumptions in $\Pi_1$: 

\begin{center} 
\AxiomC{$[A]^i$}
\noLine
\UnaryInfC{$\Sigma$}
\noLine
\UnaryInfC{$D$}
\AxiomC{$[A\supset B]^j$}
\AxiomC{$\Pi_1$}
\noLine
\UnaryInfC{$A$}
\AxiomC{$[B]^k$}
\noLine
\UnaryInfC{$\Pi_2$}
\noLine
\UnaryInfC{$C$}
\RightLabel{$_k$}
\TrinaryInfC{$C$}
\noLine
\UnaryInfC{$\Xi$}
\noLine
\UnaryInfC{$D$}
\RightLabel{$_{i, j}$}
\BinaryInfC{$D$}
\DisplayProof\qquad $\leadsto$ \qquad
\AxiomC{$\Pi_1$}
\alwaysNoLine
\UnaryInfC{$[A]$}
\UnaryInfC{$\Sigma$}
\UnaryInfC{$D$}
\UnaryInfC{$\Xi^\ast$}
\UnaryInfC{$D$}
\DisplayProof
\end{center}

\noindent $\Xi^\ast$ is constructed in the following way. Let $\rho_1\ldots \rho_n$ be the sequence of applications of rules in $\Xi$ that discharge assumptions in $\Pi_1$ from top to bottom. Let $\rho_1^\ast$ be an application of the same rule as $\rho_1$, with its major, minor and specific premises concluded as in $\rho_1$ and its arbitrary premises and conclusion replaced by $D$. If $\rho_1^\ast$ has only one arbitrary premise, conclude it with the deduction ending in the upper $D$ in the schematic representation of the reduction procedure (i.e. it is concluded by $\Sigma$, in turn the continuation of $\Pi_1$ through $A$). If $\rho_1^\ast$ has two arbitrary premises, then one is concluded as described previously, and to conclude the other, observe that in that case $\Xi$ contains a subdeduction of $D$ from the conclusion $E$ of $\rho_1$: append it to the deduction concluding the other arbitrary premise of $\rho_1$ to conclude the other arbitrary premise of $\rho_1^\ast$, deleting redundant applications of rules (i.e. those discharging assumptions that do not stand above that arbitrary premise of $\rho_1$). Continue in the same way with $\rho_2$ until you reach $\rho_n$.\bigskip 

\noindent Examples of applications of the reduction procedures will make this more concrete, and I will give two shortly. For now, notice that the reduction procedure removes the maximal formula $A\supset B$  without introducing any new ones, as it only uses (some of the) material already present in $\Xi$.\bigskip  

\noindent 5. The maximal formula is of the form $\neg A$. Convert the deduction on the left into the deduction on the right, where $\Xi^\ast$ is constructed from all those applications of rules in $\Xi$ that discharge assumptions in $\Pi_1$ as in the previous reduction procedure: 

\begin{center} 
\AxiomC{$[A]^i$}
\noLine
\UnaryInfC{$\Sigma$}
\noLine
\UnaryInfC{$C$}
\AxiomC{$[\neg A]^j$}
\AxiomC{$\Pi$}
\noLine
\UnaryInfC{$A$}
\BinaryInfC{$B$}
\noLine
\UnaryInfC{$\Xi$}
\noLine
\UnaryInfC{$C$}
\RightLabel{$_{i, j}$}
\BinaryInfC{$C$}
\DisplayProof\qquad $\leadsto$ \qquad
\AxiomC{$\Pi$}
\alwaysNoLine
\UnaryInfC{$[A]$}
\UnaryInfC{$\Sigma$}
\UnaryInfC{$C$}
\UnaryInfC{$\Xi^\ast$}
\UnaryInfC{$C$}
\DisplayProof
\end{center} 

\noindent This completes the reduction procedures for maximal formulas. 

\bigskip

\noindent \emph{Examples.} Let's illustrate the method of reusing rules that discharge assumptions in the reduced assumption with two examples. First, suppose there is a sole application of a rule below the major premise discharged by $\neg I$ and above its right arbitrary premise that discharges an assumption above the minor premise of $\neg E$, and let it be $\lor E$: 

\begin{prooftree} 
\AxiomC{$[A]^3$}
\noLine
\UnaryInfC{$\Sigma$}
\noLine
\UnaryInfC{$F$}
\AxiomC{$\Xi_1$}
\noLine
\UnaryInfC{$B\lor E$}
\AxiomC{$[\neg A]^4$}
\AxiomC{$[B]^1$}
\noLine
\UnaryInfC{$\Pi$}
\noLine
\UnaryInfC{$A$}
\BinaryInfC{$C$}
\noLine
\UnaryInfC{$\Xi_2$}
\noLine
\UnaryInfC{$D$}
\AxiomC{$[E]^2$}
\noLine
\UnaryInfC{$\Xi_3$}
\noLine
\UnaryInfC{$D$}
\RightLabel{$_{1, 2}$}
\TrinaryInfC{$D$}
\noLine
\UnaryInfC{$\Xi_4$}
\noLine
\UnaryInfC{$F$}
\RightLabel{$_{3, 4}$}
\BinaryInfC{$F$}
\end{prooftree} 

\noindent In this case any applications of rules in $\Xi_2$ can only be of $\neg E$, and any application of a rule in $\Xi_4$ that discharges assumptions (i.e. rules other than $\neg E$) discharge them in $\Xi_2$. $\Xi$ of the schematic representation of the reduction procedure for maximal formulas of the form $\neg A$ is constituted by $\Xi_1, \Xi_2, \Xi_3, \Xi_4$ in the example. $\Xi^\ast$ consists of an application of $\lor E$ with its major premise derived by $\Xi_1$, its first arbitrary premise derived by $\Pi$ and $\Sigma$, and the second arbitrary premise derived by $\Xi_3$ and applications of those rules of $\Xi_4$, call them $\Xi_4^\ast$, that do not discharge assumptions in $\Xi_2$ (these become redundant). The reduced deduction is: 

\begin{prooftree} 
\AxiomC{$\Xi_1$}
\noLine
\UnaryInfC{$B\lor E$}
\AxiomC{$[B]^1$}
\noLine
\UnaryInfC{$\Pi$}
\noLine
\UnaryInfC{$[A]$}
\noLine
\UnaryInfC{$\Sigma$}
\noLine
\UnaryInfC{$F$}
\AxiomC{$[E]^2$}
\noLine
\UnaryInfC{$\Xi_3$}
\noLine
\UnaryInfC{$[D]$}
\noLine
\UnaryInfC{$\Xi_4^\ast$}
\noLine
\UnaryInfC{$F$}
\RightLabel{$_{1, 2}$}
\TrinaryInfC{$F$}
\end{prooftree} 
	
\noindent As all discharge happens above arbitrary premises, we could also first permute the application of $\lor E$ in the initial deduction downwards to conclude with $F$. This would duplicate $\Xi_4$ and in most cases create redundant applications of rules in the copy of $\Xi_4$ under $\Xi_3$, which are to be removed. Then apply the reduction procedure. The result would be the same in this example, but in the general case this method is not enough, as applying the reduction procedure removes $\Xi_2$ from the deduction, and some of it may be needed to discharge assumptions in $\Pi$: these rules would still have to be reused somehow. 

Now consider the case in which there are more than one applications of rules discharging assumptions above the minor premise $A$ of $\neg E$ in the initial deduction, and let $\lor E$ be the top-most such rule. Then all other applications of such rules are in $\Xi_4$, and $\Xi_2$ can disappear in the reduction procedure without loss. After concluding $F$ by $\lor E$, the deduction needs to continue with those rules of $\Xi_4$ that discharge assumptions in $\Pi$. This poses no further problem: as discharge happens above arbitrary premises, we can apply the relevant rules of $\Xi_4$ with $F$ as the arbitrary premises and conclusions, deriving the major, minor, specific and remaining arbitrary premise (if any) as in $\Xi$. 

As a second example, suppose $TR$ is the top most application of a rule below the major premise discharged by $\neg I$ and above its right arbitrary premise that discharges an assumption above the minor premise of $\neg E$: 

\begin{prooftree}
\AxiomC{$[A]^3$}
\noLine
\UnaryInfC{$\Sigma$}
\noLine	
\UnaryInfC{$F$}
\AxiomC{$[\neg A]^4$}
\AxiomC{$[B]^1$}
\noLine
\UnaryInfC{$\Pi$}
\noLine
\UnaryInfC{$A$}
\BinaryInfC{$C$}
\noLine
\UnaryInfC{$\Xi_1$}
\noLine
\UnaryInfC{$D$}
\AxiomC{$[B\supset E]^2$} 
\noLine
\UnaryInfC{$\Xi_2$}
\noLine
\UnaryInfC{$D$} 
\RightLabel{$_{1, 2}$}
\BinaryInfC{$D$}
\noLine
\UnaryInfC{$\Xi_3$}
\noLine
\UnaryInfC{$F$}
\RightLabel{$_{3, 4}$}
\BinaryInfC{$F$}
\end{prooftree} 

\noindent Then $\Xi$ of the schematic representation of the reduction procedure for maximal formulas of the form $\neg A$ is constituted by $\Xi_1, \Xi_2, \Xi_3$. $\Xi^\ast$ consists of an application of $TR$ with its left arbitrary premise concluded by $\Sigma, \Pi$, and its right arbitrary premise concluded by $\Xi_2$ and those applications of $\Xi_3$, call them $\Xi_3^{\ast_1}$, needed to derive $F$ from $D$, and with its conclusion followed by applications of the rules of $\Xi_3$, call them $\Xi_3^{\ast_2}$, that discharge assumptions in $\Pi$. The reduced deduction is: 

\begin{prooftree} 
\AxiomC{$[B]^1$}
\noLine
\UnaryInfC{$\Pi$}
\noLine
\UnaryInfC{$[A]$}
\noLine
\UnaryInfC{$\Sigma$}
\noLine	
\UnaryInfC{$F$}
\AxiomC{$[B\supset E]^2$} 
\noLine
\UnaryInfC{$\Xi_2$}
\noLine
\UnaryInfC{$[D]$} 
\noLine
\UnaryInfC{$\Xi_3^{\ast_1}$}
\noLine
\UnaryInfC{$F$}
\RightLabel{$_{1, 2}$}
\BinaryInfC{$F$}
\noLine
\UnaryInfC{$\Xi_3^{\ast_2}$}
\end{prooftree} 
 
\noindent Once more, as all discharge happens above arbitrary premises, constructing $\Xi_3^{\ast_2}$ from $\Xi_3$ poses no further problem. 

\bigskip

\noindent\textbf{C} has 10 rules, 4 of which are elimination rules. So there are 40 permutative reduction procedures. Most of them are handled by permuting the application of the elimination rule upwards. The 4 cases where the major premise of an elimination rule is concluded by $\neg E$ do not really involve any permutation upwards, but I will class them with the permutative reduction procedures. None of this presents any difficulty, so I will only give some of the permutative reduction procedures as examples, the others being similar. 

\bigskip 

\noindent \emph{Premutative Reduction Procedures for Maximal Segments} 

\noindent 1. The major premise of $\supset E$ is derived by $\lor I$. Convert the deduction on top into the deduction below: 

\begin{center}
\AxiomC{$\Pi_1$}
\noLine
\UnaryInfC{$A$}
\AxiomC{$[A\lor B]^i$}
\noLine
\UnaryInfC{$\Pi_2$}
\noLine
\UnaryInfC{$C\supset D$}
\RightLabel{$_i$}
\BinaryInfC{$C\supset D$}
\AxiomC{$\Sigma_1$}
\noLine
\UnaryInfC{$C$}
\AxiomC{$[D]^j$}
\noLine
\UnaryInfC{$\Sigma_2$}
\noLine
\UnaryInfC{$E$}
\RightLabel{$_j$}
\TrinaryInfC{$E$}
\DisplayProof\bigskip

$\leadsto$\bigskip

\AxiomC{$\Pi_1$}
\noLine
\UnaryInfC{$A$}
\AxiomC{$[A\lor B]^i$}
\noLine
\UnaryInfC{$\Pi_2$}
\noLine
\UnaryInfC{$C\supset D$}
\AxiomC{$\Sigma_1$}
\noLine
\UnaryInfC{$C$}
\AxiomC{$[D]^j$}
\noLine
\UnaryInfC{$\Sigma_2$}
\noLine
\UnaryInfC{$E$}
\RightLabel{$_j$}
\TrinaryInfC{$E$}
\RightLabel{$_i$}
\BinaryInfC{$E$}
\DisplayProof
\end{center} 

\noindent 2. The major premise of $\supset E$ is derived by $\supset I$. Convert the deduction on top into the deduction below: 

\begin{center} 
\AxiomC{$\Pi_1$}
\noLine
\UnaryInfC{$B$}
\AxiomC{$[A\supset B]^i$}
\noLine
\UnaryInfC{$\Pi_2$}
\noLine
\UnaryInfC{$C\supset D$}
\RightLabel{$_i$}
\BinaryInfC{$C\supset D$}
\AxiomC{$\Sigma_1$}
\noLine
\UnaryInfC{$C$}
\AxiomC{$[D]^j$}
\noLine
\UnaryInfC{$\Sigma_2$}
\noLine
\UnaryInfC{$E$}
\RightLabel{$_j$}
\TrinaryInfC{$E$}
\DisplayProof\bigskip

$\leadsto$\bigskip

\AxiomC{$\Pi_1$}
\noLine
\UnaryInfC{$B$}
\AxiomC{$[A\supset B]^i$}
\noLine
\UnaryInfC{$\Pi_2$}
\noLine
\UnaryInfC{$C\supset D$}
\AxiomC{$\Sigma_1$}
\noLine
\UnaryInfC{$C$}
\AxiomC{$[D]^j$}
\noLine
\UnaryInfC{$\Sigma_2$}
\noLine
\UnaryInfC{$E$}
\RightLabel{$_j$}
\TrinaryInfC{$E$}
\RightLabel{$_i$}
\BinaryInfC{$E$}
\DisplayProof
\end{center} 

\noindent 3. The major premise of $\supset E$ is derived by $\land E$. Convert the deduction on top into the deduction below: 

\begin{center} 
\AxiomC{$\Pi_1$}
\noLine
\UnaryInfC{$A\land B$}
\AxiomC{$[A]^i \ [B]^j$}
\noLine
\UnaryInfC{$\Pi_2$}
\noLine
\UnaryInfC{$C\supset D$}
\RightLabel{$_{i, j}$}
\BinaryInfC{$C\supset D$}
\AxiomC{$\Sigma_1$}
\noLine
\UnaryInfC{$C$}
\AxiomC{$[D]^k$}
\noLine
\UnaryInfC{$\Sigma_2$}
\noLine
\UnaryInfC{$E$}
\RightLabel{$_k$}
\TrinaryInfC{$E$}
\DisplayProof\bigskip

$\leadsto$\bigskip 

\AxiomC{$\Pi_1$}
\noLine
\UnaryInfC{$A\land B$}
\AxiomC{$[A]^i \ [B]^j$}
\noLine
\UnaryInfC{$\Pi_2$}
\noLine
\UnaryInfC{$C\supset D$}
\AxiomC{$\Sigma_1$}
\noLine
\UnaryInfC{$C$}
\AxiomC{$[D]^k$}
\noLine
\UnaryInfC{$\Sigma_2$}
\noLine
\UnaryInfC{$E$}
\RightLabel{$_k$}
\TrinaryInfC{$E$}
\RightLabel{$_{i, j}$}
\BinaryInfC{$E$}
\DisplayProof
\end{center} 

\bigskip 

\noindent 4. The major premise of $\neg E$ is derived by $TR$. Convert the deduction to the left of $\leadsto$ into the one on its right: 

\begin{center} 
\AxiomC{$[A]^i$}
\noLine
\UnaryInfC{$\Pi_1$}
\noLine
\UnaryInfC{$\neg C$}
\AxiomC{$[A\supset B]^j$}
\noLine
\UnaryInfC{$\Pi_2$}
\noLine
\UnaryInfC{$\neg C$}
\RightLabel{$_{i, j}$}
\BinaryInfC{$\neg C$}
\AxiomC{$\Sigma_1$}
\noLine
\UnaryInfC{$C$}
\BinaryInfC{$D$}
\DisplayProof$\qquad\leadsto\qquad $
\AxiomC{$[A]^i$}
\noLine
\UnaryInfC{$\Pi_1$}
\noLine
\UnaryInfC{$\neg C$}
\AxiomC{$\Sigma_1$}
\noLine
\UnaryInfC{$C$}
\BinaryInfC{$D$}
\AxiomC{$[A\supset B]^j$}
\noLine
\UnaryInfC{$\Pi_2$}
\noLine
\UnaryInfC{$\neg C$}
\AxiomC{$\Sigma_1$}
\noLine
\UnaryInfC{$C$}
\BinaryInfC{$D$}
\RightLabel{$_{i, j}$}
\BinaryInfC{$D$}
\DisplayProof
\end{center} 

\noindent 5. In the case a major premise of an elimination rule is concluded by $\neg E$, remove the application of the elimination rule and use $\neg E$ to conclude the discharged assumption(s) of (one of) the side deductions concluding with an arbitrary premise instead. For instance, convert the deduction to the left of $\leadsto$ into the one on its right: 

\begin{center}
\AxiomC{$\Pi_1$}
\noLine
\UnaryInfC{$\neg A$}
\AxiomC{$\Pi_2$}
\noLine
\UnaryInfC{$A$}
\BinaryInfC{$B\lor C$}
\AxiomC{$[B]^1$}
\noLine
\UnaryInfC{$\Sigma_1$}
\noLine
\UnaryInfC{$D$}
\AxiomC{$[C]^2$}
\noLine
\UnaryInfC{$\Sigma_1$}
\noLine
\UnaryInfC{$D$}
\RightLabel{$_{1, 2}$}
\TrinaryInfC{$D$}
\DisplayProof
$\qquad\leadsto\qquad $
\AxiomC{$\Pi_1$}
\noLine
\UnaryInfC{$\neg A$}
\AxiomC{$\Pi_2$}
\noLine
\UnaryInfC{$A$}
\BinaryInfC{$[B]$}
\noLine
\UnaryInfC{$\Sigma_1$}
\noLine
\UnaryInfC{$D$}
\DisplayProof 
\end{center}

\noindent Alternatively, we could have concluded $C$ by$\neg E$ and used $\Sigma_2$ to conclude $D$. The normal form of deductions is therefore not unique. 

\bigskip

\noindent This completes the permutative reduction procedures. 

\bigskip

\noindent Repeated application of a permutative reduction procedure reduces the length of a maximal segment by permuting applications of elimination rules upwards in the deduction. As noted earlier, the first formula of a segment can only be one discharged by an introduction rule, and so repeated application of a permutative reduction procedure turns a maximal segment into a maximal formula. At the top of every maximal segment, there stands a maximal formula, so to speak. 

\begin{definition}[Rank of Deductions]
\normalfont The \emph{rank} of a deduction $\Pi$ is the pair $\langle d, l\rangle$, where $d$ is the highest degree of a maximal formula or maximal segment in $\Pi$ or $0$ if there is none, and $l$ is the sum of sum of the lengths of maximal segments of highest degree and number of maximal formulas in $\Pi$. $\langle d, l\rangle < \langle d', l'\rangle$ iff either (i) $d<d'$ or (ii) $d=d'$ and $l<l'$.
\end{definition} 

\noindent Applying reduction procedures for maximal formulas cannot introduce new maximal formulas into the reduced deduction, but it may increase the lengths of maximal segments that were in the initial deduction.\footnote{It may also shorten maximal segments, i.e. if the arbitrary premises marked by $C$ or $D$ in the reduction procedures form part of one.} In the case of maximal formulas of form $A\land B$, this can happen if $A$ is concluded by a rule in $\Sigma_1$ or $B$ is in $\Sigma_2$ (i.e. $\Sigma_1$ or $\Sigma_2$ is not empty) and some formula occurrence in the assumption class to which the formulas discharged by $\land E$ belong is the major premise of an elimination rule in $\Pi_1$. Similarly for maximal formulas of the form $A\lor B$. In the case of maximal formulas of the form $A\supset B$, this can happen in reduction procedure 3 if $B$ is concluded by a rule in $\Sigma$ (i.e. $\Sigma$ is not empty) and some formula occurrence in the assumption class to which the formulas discharged by $\supset E$ belong is the major premise of an elimination rule in $\Pi_2$, and in reduction procedure 4 if $A$ is concluded by a rule in $\Pi_1$ (i.e. $\Pi_1$ is not empty) and some formula occurrence in the assumption class to which the minor assumptions discharged by $TR$ belong is the major premises of an elimination rule in $\Sigma$. The case of maximal formulas of the form $\neg A$ is similar to the last one with $TR$. 

Any maximal segment that suffers an increase in length as a result of a reduction procedure is, however, of lower degree than the maximal formula removed, as the formula that form part of the segment are subformulas of the latter. Hence applying a reduction procedure for maximal formulas cannot increase the rank of a deduction.  

An application of a permutative reduction procedure reduces the length of a maximal segment by 1 or removes it entirely, if the maximal segment is of length 1 (i.e. it is the conclusion of $\neg E$ and major premise of an elimination rule). In the latter case, another maximal segment is also reduced by 1, namely should the conclusion $D$ in the schematic representation of the procedure above form part of a maximal segment. The other permutative reduction procedures may increase the lengths of maximal segments that were in the initial deduction. To ensure the decrease of the rank of a deduction, the permutative reduction procedures must be applied with a strategy. 
  
Adopting the convention that a deduction that already is in normal form converts into itself, we have: 

\begin{theorem}\label{normalisation}
Any deduction in \textbf{C} can be converted into a deduction in normal form.
\end{theorem}

\noindent\emph{Proof.} By Lemma 1, it suffices to consider deductions in which introduction rules discharge exactly one major assumption. The theorem follows by the considerations of the paragraphs immediately preceding the theorem and an induction over the rank of deductions. Applying reduction procedures for maximal formulas cannot increase the rank of a deduction, and as a maximal formula is removed, applying a reduction procedure to a maximal formula of highest degree decreases the rank of the deduction. Permutative reduction procedures must be applied so as to avoid an increase of a length of segments of highest degree. This can be achieved by applying one to a maximal segment of highest degree such that no maximal segment of highest degree stands above it in the deduction. This reduces the rank of the deduction.\bigskip

\noindent The deduction in normal form has the same undischarged assumptions as the initial deduction:

\begin{theorem}\label{sameassumptions}
For any deduction in \textbf{C}, there is a deduction in normal form of the same conclusion from the same undischarged assumptions.
\end{theorem}
\noindent \emph{Proof.} By theorem \ref{normalisation} and the ban on vacuous discharge.\bigskip

\noindent This is a noteworthy difference to normalisation in Gentzen's and Prawitz's systems of classical intuitionist logic, where during the course of the normalisation procedure typically some undischarged assumptions of the deduction are removed together with maximal formulas above which they occur. 

The form of normal deductions has other noteworthy features: 

\begin{theorem}\label{majorpremisesassumptions}
If $\Pi$ is a deduction in normal form, then all major premises of elimination rules are (discharged or undischarged) assumptions of $\Pi$.
\end{theorem} 

\noindent\emph{Proof.} This is a consequence of the way the permutative reduction procedures are applied in normalisation.\bigskip 

\noindent Deductions in normal form in \textbf{C} are particularly perspicuous. 

\begin{definition}[Subformula Property]
\normalfont A deduction $\Pi$ of a conclusion $C$ from the undischarged assumptions $A_1\ldots A_n$ has the \emph{subformula property} if every formula on the deduction is a subformula either of $C$ or of $A_1\ldots A_n$. 
\end{definition}

\noindent The crucial detail of deductions in normal form in \textbf{C} that ensures that they enjoy the subformula property is that the major premise $\neg A$ of $\neg E$ is not the major assumption discharged by $\neg I$, and because the minor assumption discharged by $TR$ is a subformula of the major assumption, which is not the major premise of $\supset E$. To prove the subformula property formally, we need the notion of a branch:  

\begin{definition}[Branch]
\normalfont A \emph{branch} in a deduction is a sequence of formula occurrences $\sigma_1\ldots \sigma_n$ such that $\sigma_1$ is an assumption of the deduction that is neither discharged by an elimination rule nor the major assumption discharged by an introduction rule other than $\neg I$ or $TR$, $\sigma_n$ is either the conclusion of the deduction or the minor premise of $\supset E$ or $\neg E$, and for each $n>i$: 

\noindent (i) if $\sigma_i$ is the major premise of an elimination rule other than $\neg E$, $\sigma_{i+1}$ is an assumption discharged by it, and if it is the major premise of $\neg E$, $\sigma_{i+1}$ is the conclusion of the rule; 

\noindent (ii) if $\sigma_i$ is the specific premise of an introduction rule, $\sigma_{i+1}$ is a major assumption discharged by it; 

\noindent (iii) and if $\sigma_i$ is an arbitrary premise (of an introduction or an elimination rule rule), $\sigma_{i+1}$ is the conclusion of the rule. 
\end{definition} 

\begin{definition}[Order of Branches]
\normalfont A branch is of order $0$ if its last formula is the conclusion of the deduction; it is of order $n+1$ if its last formula is the minor premise of $\supset E$ or of $\neg E$ such that its major premise is on a branch of order $n$. 
\end{definition} 

\noindent A branch of order $0$ is also called a \emph{main branch} in the deduction. 

\begin{corollary} 
If any major premises of elimination rules are on a branch in a deduction in normal form, then they precede any major assumptions discharged by introduction rules that are on the branch.
\end{corollary} 

\noindent \emph{Proof.} By theorem \ref{majorpremisesassumptions}, the major premises of elimination rules that occur on a branch in a deduction in normal form are assumptions. Hence they are not the last formulas of any segments, and in particular they are not the last formulas of any segments beginning with discharged major assumptions of introduction rules.\bigskip

\noindent $TR$ and $\neg I$ are introduction rules without specific premises. Branches move from their premises to their conclusions. A branch in a deduction in \textbf{C} begins with an undischarged assumption or an assumption discharged by $\neg I$ or $TR$. If an assumption discharged by $TR$ or $\neg I$ was also an arbitrary premise of the rule, there would be vacuous discharge above the other arbitrary premises. But vacuous discharge was banned. Hence the assumptions discharged by $TR$ and $\neg I$ can only be major, minor or specific premises of rules. In deductions in normal form, the first option does not occur in case of the major assumptions discharged. Hence the major assumptions discharged by $TR$ and $\neg I$ are minor premises of $\neg E$, $\supset E$ or specific premises of $\supset I$ or $\lor I$, and consequently subformulas of other formulas on the deduction. The minor premises discharged by $TR$ and $\neg E$ are subformulas of the major premises discharged. This, in a nutshell, guarantees that deductions in normal form enjoy the subformula property. 

A branch in a deduction in normal form in \textbf{C} can be partitioned into an \emph{E-part}, the first part of the branch, which consists of major premises of elimination rules, and an \emph{I-part}, which consists of sequences of segments that are the major assumptions discharged by introduction rules. Separating the two parts is the \emph{minimal formula} or \emph{minimal segment}. Either part may be empty: some branches in normal deductions consist of only an E-part, some of only an I-part, and in the case of a deduction that consists of a single formula $A$, both parts are empty and there is only a minimal formula. 

For brevity we may speak of a segment being the premise, conclusion or discharged assumption of the rule of which its last or first formula is the premise, conclusion or discharged assumption. 

An induction over the order of branches establishes the following result: 

\begin{corollary}\label{subformula}
Deductions in normal form have the subformula property.
\end{corollary} 

\noindent \emph{Proof.} By inspection of the rules and an induction over the order of branches. Consider a branch of order 0. If it has an E-part, it begins with a sequence of formulas and segments that are major premises of elimination rules, going from major premise to assumption discharged by the elimination rule, until it reaches a specific formula of an introduction rule, and then continues with segments discharged by introduction rules, until it reaches the conclusion of the deduction. All formulas in the latter part of the branch are subformulas of the conclusion of the deduction. All the formulas on the former part of the branch are subformulas of an assumption that remains undischarged in the deduction. If the branch does not have an E-part, it begins with specific premises of introduction rules. This is because the major assumptions discharged by $TR$ and $\neg E$ form branches consisting of only a minimal formula, because due to the normal form of the deduction and the ban on vacuous discharge, they can only be the specific premises of introduction rules. In a deduction in normal form branches beginning with the major assumptions discharged by $TR$ and $\neg I$ cannot contain any major premises of elimination rules. Hence the major assumptions discharged by $TR$ and $\neg I$ are subformulas of the final formula of the branch, and therefore so are the minor premises discharged by these rules. The only tricky case to consider is $\neg E$: here the branch goes from the major premise of the form $\neg A$ to a conclusion $C$, which may not share any subformulas. But the case is really no different from applications of $\bot E$ in Gentzen's and Prawitz's system of natural deduction for intuitionist logic: the crucial point is that $C$ is not the major premise of an elimination rule, and so it is on the I-part of the branch, and hence a subformula of the last formula of the branch. Any conclusion of $\neg E$ on a branch of a deduction in normal form is either the last formula of the branch, or the specific premise of an introduction rule, or the first formula of a segment ending in a specific premise of an introduction rule, and hence it is a subformula of the conclusion of the deduction. This completes the basis of the induction. A branch that ends in the minor premise of $\supset E$ or $\neg E$ ends in a formula that is a subformula of a branch of lower order, and hence the theorem holds by induction over the order of branches.\bigskip 

\noindent As a corollary of this corollary we have: 

\begin{corollary}
For any deduction in \textbf{C}, there is a deduction of the same conclusion from the same undischarged assumptions with the subformula property. 
\end{corollary} 

\noindent\emph{Proof.} From theorem \ref{sameassumptions} and corollary \ref{subformula}.\bigskip

\noindent Finally, normalisation yields a direct proof of the consistency of \textbf{C}. Let a \emph{proof} in \textbf{C} be a deduction that has no undischarged assumptions. 

\begin{corollary}
If there is a proof of $A$ in \textbf{C}, then there is one that ends with an introduction rule.
\end{corollary}

\noindent \emph{Proof.} Elimination rules do not discharge assumptions above their major premises. Hence if in a deduction in normal form there is a main branch that begins with the major premises of an elimination rule and does not have an I-part, it is not a proof. Contraposing and applying theorem \ref{normalisation}, if there is a proof of $A$ in \textbf{C}, then there is one that ends with an introduction rule. 

\begin{corollary}
\textbf{C} is consistent. 
\end{corollary}

\noindent\emph{Proof.} Suppose there is a proof of an atomic formula $p$ in \textbf{C}. Then there is a proof in normal form of $p$. But as $p$ contains no connective, it cannot end in an introduction rule, and hence it is not a proof. Contradiction.\bigskip 

\noindent A similar argument establishes that by the form of normal deductions, a proof of $A\land \neg A$ would end with an introduction rule, hence both $A$ and $\neg A$ would need to have been derived from no assumptions, which is impossible.

\section{Adding the Quantifiers} 
To close this paper, let's consider what happens when \textbf{C} is extended to quantificational logic. The language is extended in standard fashion to contain the connectives $\forall$, $\exists$, constants, function and predicate symbols and variables. The language has two disjoint sets of variables, the \emph{parameters} $a, b, c\ldots$ playing the role of free variables, and the variables to be bound by the quantifiers $x, y, z\ldots$, which do not occur free in formulas. The terms of the language are built up from the parameters, constant symbols and function symbols. We may call an expression that is like a formula or a term, but containing free variables instead of parameters, a pseudo-formula or a pseudo-term. 

Let $A_x^a$ be the result of substituting the variable $x$ for the parameter $a$ in $A$. Let $A^x_t$ be the result of substituting all occurrences of the variable $x$ in $A$ by the term $t$, where it is assumed that $t$ is free for $x$ in $A$, i.e. no variable in the pseudo-term $t$ becomes bound as a result of the substitution. 

The usual elimination rule for the existential quantifier already has the form of general elimination rules. The general elimination rule for the universal quantifier has the same form with a different use of terms: 

\begin{center} 
\AxiomC{$\Sigma$}
\noLine
\UnaryInfC{$\exists x A$}
\AxiomC{$[A^x_a]^i$}
\noLine
\UnaryInfC{$\Pi$}
\noLine
\UnaryInfC{$C$}
\RightLabel{$_{\exists E \ i}$}
\BinaryInfC{$C$}
\DisplayProof\qquad \qquad
\AxiomC{$\Sigma$}
\noLine
\UnaryInfC{$\forall x A$}
\AxiomC{$[A^x_t]^i$}
\noLine
\UnaryInfC{$\Pi$}
\noLine
\UnaryInfC{$C$}
\RightLabel{$_{\forall E \ i}$}
\BinaryInfC{$C$}
\DisplayProof
\end{center} 

\noindent where in $\exists E$, parameter $a$ does not occur in $\exists xA$, nor in $C$, nor in any formulas undischarged in $\Pi$ except those of the assumption class $[A^x_a]$. 

The following are the general introduction rules for the quantifiers:

\begin{center} 
\AxiomC{$\Sigma$}
\noLine
\UnaryInfC{$A^x_t$}
\AxiomC{$[\exists xA]^i$}
\noLine
\UnaryInfC{$\Pi$}
\noLine
\UnaryInfC{$C$}
\RightLabel{$_{\exists I \ i}$}
\BinaryInfC{$C$} 
\DisplayProof\qquad \qquad
\AxiomC{$\Sigma$}
\noLine
\UnaryInfC{$A^x_a$}
\AxiomC{$[\forall x A]^i$}
\noLine
\UnaryInfC{$\Pi$}
\noLine
\UnaryInfC{$C$}
\RightLabel{$_{\forall I \ i}$}
\BinaryInfC{$C$}
\DisplayProof
\end{center} 

\noindent where in $\forall I$, parameter $a$ does not occur in any undischarged assumption of $\Sigma$.\footnote{The general introduction rule for $\exists$ is Milne's \citep{milneinversion}.}  
 
It is worth remarking that the rules for both quantifiers have the same form and differ only with respect to the occurrences of terms and parameters and consequently where restrictions on parameters are imposed. 

Milne observes that adding rules for both quantifiers upsets the subformula. It is instructive to see why. Consider the classically, but not intuitionistically, valid inference of $\forall x(Fx\lor A)\vdash\forall x Fx \lor A$ ($x$ not free in $A$). It may be derived in the following way:

\begin{prooftree} 
\AxiomC{$[A]^4$} 
\UnaryInfC{$\forall xFx\lor A$} 
\AxiomC{$\forall x(Fx\lor A)$} 
\UnaryInfC{$F_a^x\lor A$}
\AxiomC{$[F_a^x]^1$} 	
\AxiomC{$[\neg A]^5$}
\AxiomC{$[A]^2$}
\BinaryInfC{$F_a^x$}
\RightLabel{$_{1, 2}$}
\TrinaryInfC{$F_a^x$}
\AxiomC{$[\forall x Fx]^3$} 
\RightLabel{$_3$}
\BinaryInfC{$\forall x Fx$} 
\UnaryInfC{$\forall xFx\lor A$} 
\RightLabel{$_{4, 5}$}
\BinaryInfC{$\forall xFx\lor A$} 
\end{prooftree} 

\noindent $\neg A$ is a maximal formula. An attempt to remove it from the deduction by applying the reduction procedure for maximal formulas of this form gives the following: 

\begin{prooftree}
\AxiomC{$\forall x(Fx\lor A)$} 
\UnaryInfC{$F_a^x\lor A$}
\AxiomC{$[F_a^x]^1$}
\AxiomC{$[\forall x Fx]^3$} 
\RightLabel{$_3$}
\BinaryInfC{$\forall x Fx$} 
\UnaryInfC{$\forall xFx\lor A$} 
\AxiomC{$[A]^2$} 
\UnaryInfC{$\forall xFx\lor A$} 
\RightLabel{$_{1, 2}$}
\TrinaryInfC{$\forall xFx\lor A$}
\end{prooftree} 

\noindent The application of $\forall I$ is now incorrect. 

The problem is that $\Xi$ in the schematic representation of the reduction procedure for maximal formulas of the form $\neg A$ may contain applications of rules that discharge formulas in which parameters of applications of $\forall I$ lower down in the deduction occur, but in $\Xi^\ast$, these applications become incorrect as they are permuted upwards. 

So, for instance in the first example of an application for the reduction procedure for maximal formulas of the form $\neg A$ given in the previous section, there may be an application of $\forall I$ in $\Xi_4$ that is indispensable for the deduction of $F$ from $D$ (i.e. if they are different), where $E$ contains the parameter of that application: this application becomes incorrect in $\Xi_4^\ast$. 

A similar problem occurs in case the maximal formula is the major assumption discharged by $TR$. 

The situation is different with applications of $\exists E$, as there the restriction on parameters is imposed on a deduction above an arbitrary premise. If there is an application of $\exists E$ in $\Xi_4$, we apply it after the application of $\lor E$, and the problem does not arise. 

Let $\mathbf{C^\exists}$ be \textbf{C} plus the rules for the existential quantifier. We will give reduction procedures for maximal formulas of the form $\exists xA$ and prove a normalisation theorem for this system. 

As we have an unlimited amount of parameters at our disposal, we may adopt the convention that every application of $\exists E$ has its own parameter, so that the parameter of an application of $\exists E$ occurs only in its discharged assumption and formulas derived from it. Consequently, the parameter occurs only above the application of the rule in a deduction, and any application of $\exists E$ below it has a different parameter. Call this the \emph{parameter convention}.\footnote{Alternatively we could assume the parameters in deductions are renamed as part of the reduction procedure, wherever necessary.}

Inspection of the reduction procedures for the propositional connectives shows that, if the parameter convention is upheld, then any correct application of $\exists E$ in the initial deduction remains correct in the reduced deduction. The same holds for the following reduction procedure for maximal formulas of the form $\exists x A$, where $\Xi_t^a$ is the result of substituting the term $t$ for the parameter $a$ throughout $\Xi$. We add to the list of reduction procedures of the previous section:\bigskip

\noindent 6. The maxima formula is of the form $\exists xA$. Convert the deduction on the left into the deduction on the right: 

\begin{center} 
\AxiomC{$\Sigma$}
\noLine
\UnaryInfC{$A^x_t$}
\AxiomC{$[\exists x A]^i$}
\AxiomC{$[A^x_a]^j$}
\noLine
\UnaryInfC{$\Xi$}
\noLine
\UnaryInfC{$C$}
\RightLabel{$_j$}
\BinaryInfC{$C$}
\noLine
\UnaryInfC{$\Pi$} 
\noLine
\UnaryInfC{$D$}
\RightLabel{$_i$}
\BinaryInfC{$D$} 
\DisplayProof\quad $\leadsto$\quad
\AxiomC{$\Sigma$}
\noLine
\UnaryInfC{$[A^x_t]$}
\noLine
\UnaryInfC{$\Xi ^a_t$}
\noLine
\UnaryInfC{$C$}
\noLine
\UnaryInfC{$\Pi$} 
\noLine
\UnaryInfC{$D$}
\DisplayProof
\end{center} 

\noindent The additional permutative reduction procedures are evident and pose no further difficulty. Neither does extending the proof of theorem \ref{normalisation} to $\mathbf{C^\exists}$, and we have: 

\begin{theorem}\label{normalisation2}
Deductions in $\mathbf{C^\exists}$ normalise. 
\end{theorem}

\noindent Consequently, for the standard notion of a subformula applicable to quantificational logic, where all substitution instances of a formula $\exists x A$ are counted amongst its subformulas, we have: 

\begin{corollary}\label{subformula2}
Deductions in normal form $\mathbf{C^\exists}$ have the subformula property.
\end{corollary} 

\noindent Finally, as vacuous discharge remains banned in $\mathbf{C^\exists}$, we have: 

\begin{corollary}
For any deduction in $\mathbf{C^\exists}$, there is a deduction of the same conclusion from the same undischarged assumptions with the subformula property. 
\end{corollary} 

\noindent\emph{Proof.} From theorem \ref{normalisation2} and corollary \ref{subformula2}.

\section{Conclusion} 
Having adopted the unique discharge convention meant that, for the purposes of normalisation, consideration was restricted to applications of general introduction rules that discharge exactly one major assumption. After normalisation, it would be possible to collapse certain sequences of applications of general introduction rules into one. On the other hand, the unique discharge convention allows for an easy transposition of \textbf{C} into a more conventional system. The conclusion of an application of an introduction rule in Gentzen's system obviously occurs exactly once in a deduction, so if the unique discharge convention is upheld, there is a simple correspondence between deductions in Gentzen's system and in the present system with general introduction rules. Instead of assuming, concluding and discharging the major assumption $A\ast B$, we only conclude it, without making and discharging the assumption. This works for $\land$, $\lor I$, $\supset I$ and $\exists I$, but of course not for $\neg I$ and $TR$: them we must leave as they are. Furthermore, although it is not necessary for a complete system of classical logic to allow the discharge of an assumption $A$ above $B$ in $\supset I$, as this rule is derivable from $\supset I$ and $TR$, doing so does not upset the subformula property. The following is therefore a complete system of classical quantificational logic in which for every deduction, there is a deduction that satisfies the subformula property: 

\begin{center}
\bottomAlignProof
\AxiomC{$A$} 
\AxiomC{$B$}
\LeftLabel{$\land I$: \ }
\BinaryInfC{$A\land  B$}
\DisplayProof\qquad\qquad
\bottomAlignProof
\AxiomC{$A\land B$}
\AxiomC{$[A]^i \ [B]^j$}
\noLine
\UnaryInfC{$\Pi$}
\noLine
\UnaryInfC{$C$}
\RightLabel{$_{i, j}$}
\LeftLabel{$\land E$: \ }
\BinaryInfC{$C$}
\DisplayProof
	
\bigskip

\bottomAlignProof
\AxiomC{$[A]^i$}
\noLine
\UnaryInfC{$\Pi$}
\noLine
\UnaryInfC{$B$}
\LeftLabel{$\supset I$: \ }
\RightLabel{$_i$}
\UnaryInfC{$A\supset B$}
\DisplayProof\qquad\qquad
\bottomAlignProof
\AxiomC{$A\supset B$}
\AxiomC{$A$}
\AxiomC{$[B]^i$}
\noLine
\UnaryInfC{$\Pi$}
\noLine
\UnaryInfC{$C$}
\RightLabel{$_i$}
\LeftLabel{$\supset E$: \ }
\TrinaryInfC{$C$}
\DisplayProof

\bigskip

\AxiomC{$[A]^i$}
\noLine
\UnaryInfC{$\Pi$}
\noLine
\UnaryInfC{$C$}
\AxiomC{$[A\supset B]^j$}
\noLine
\UnaryInfC{$\Sigma$}
\noLine
\UnaryInfC{$C$}
\LeftLabel{$TR$: \ }
\RightLabel{$_{i, j}$}
\BinaryInfC{$C$}
\DisplayProof

\bigskip

\bottomAlignProof
\AxiomC{$A$}
\LeftLabel{$\lor I$: \ }
\UnaryInfC{$A\lor B$}
\DisplayProof\quad
\bottomAlignProof
\AxiomC{$B$}
\UnaryInfC{$A\lor B$}
\DisplayProof\qquad\qquad
\bottomAlignProof
\AxiomC{$A\lor B$}
\AxiomC{$[A]^i$}
\noLine
\UnaryInfC{$\Pi$}
\noLine
\UnaryInfC{$C$}
\AxiomC{$[B]^j$}
\noLine
\UnaryInfC{$\Sigma$}
\noLine
\UnaryInfC{$C$}
\LeftLabel{$\lor E$: \ }
\RightLabel{$_{i, j}$}
\TrinaryInfC{$C$}
\DisplayProof
	
\bigskip

\bottomAlignProof
\AxiomC{$[A]^i$}
\noLine
\UnaryInfC{$\Pi$}
\noLine
\UnaryInfC{$C$}
\AxiomC{$[\neg A]^j$}
\noLine
\UnaryInfC{$\Sigma$}
\noLine
\UnaryInfC{$C$}
\RightLabel{$_{i, j}$}
\LeftLabel{$\neg I$: \ }
\BinaryInfC{$C$}
\DisplayProof\qquad\qquad
\bottomAlignProof
\AxiomC{$\neg A$}
\AxiomC{$A$}
\LeftLabel{$\neg E$: \ }
\BinaryInfC{$C$}
\DisplayProof

\bigskip 

\bottomAlignProof
\AxiomC{$A^x_t$}
\LeftLabel{$\exists I$: \ } 
\UnaryInfC{$\exists xA$}
\DisplayProof\qquad\qquad
\bottomAlignProof
\AxiomC{$\exists x A$}
\AxiomC{$[A^x_a]^i$}
\noLine
\UnaryInfC{$\Pi$}
\noLine
\UnaryInfC{$C$}
\LeftLabel{$\exists E$: \ } 
\RightLabel{$_i$}
\BinaryInfC{$C$}
\DisplayProof

\end{center}

\noindent Finally, we could also define $\neg$ in terms of $\bot$ and replace the two negation rules with one rule: 

\begin{center}
\AxiomC{$\bot$}
\LeftLabel{$\bot E$: \ } 
\UnaryInfC{$C$}
\DisplayProof
\end{center} 

\noindent This rule can be restricted to atomic conclusions, and certain economies ensue in the number of reduction procedures required to prove normalisation.

\bigskip

\setlength{\bibsep}{0pt}
\bibliographystyle{chicago}
\bibliography{KurbisGenIntroRulesCL}
\end{document}

\title{Correction regarding `Normalisation and Subformula Property for a System of Classical Logic with Tarski's Rule'}
\author{Nils K\"urbis}
\date{}
\maketitle

%\begin{center}
%Published in \emph{Archive for Mathematical Logic} \\ 
%\url{}\bigskip
%\end{center}

\noindent My enthusiasm for Milne's system of classical logic with general introduction and elimination rules \citep{milneinversion,milnesubformula} had got the better of me when I announced, in Theorem 2 of \citep{kurbisgenrulescl}, that for any deduction in \textbf{C}, there is a deduction in normal form of the same conclusion from the same undischarged assumptions. The `proof' claims that this was due to the ban on vacuous discharge. This, however, is of course mistaken. The possibility of banning vacuous discharge from Milne's system is noteworthy, but unfortunately lacks any such miraculous effect: evidently the reduction procedures for maximal formulas remove entire parts of deductions, vacuous discharge or not, and any undischarged assumptions in such parts that are not also in assumption classes in parts that remain will no longer be undischarged assumptions of the deduction in normal form. Consequently, Corollary 3, a corollary to Theorem 2, is in error on too: it is incorrect that for any deduction in \textbf{C}, there is a deduction of the same conclusion from the same undischarged assumptions with the subformula property. Similarly for Corollary 7, where the corresponding mistaken claim is made for the system of quantificational logic with general introduction and elimination rules for the existential quantifier. Whether this mistake is not just embarrassing, and I had in mind another result concerning the form of deductions in normal form that depends on the ban of vacuous discharge, I was not in position to reconstruct anymore from any notes on this article that remain.

\bigskip

\setlength{\bibsep}{0pt}
\bibliographystyle{chicago}
\bibliography{KurbisGenIntroRulesCLCorrection}